\documentstyle[12pt]{article}
\newcommand{\be}{\begin{equation}}
\newcommand{\ee}{\end{equation}}
\newcommand{\ba}{\begin{eqnarray}}
\newcommand{\ea}{\end{eqnarray}}

\begin{document}
\setlength{\baselineskip}{.7cm}
\renewcommand{\thefootnote}{\fnsymbol{footnote}}
\sloppy
 
\begin{center}
{\Large \bf  Linear stochastic dynamics\\ with nonlinear fractal
properties}
 
\vskip .17in
Didier Sornette$^{1,2}$
 
{\it $^1$ Laboratoire de Physique de la Mati\`ere Condens\'ee, CNRS
URA190\\ Universit\'e des Sciences, B.P. 70, Parc Valrose, 06108 Nice Cedex
2,
France \\
 
$^2$ Department of Earth and Space Science\\ and Institute of Geophysics
and
Planetary Physics\\ University of California, Los Angeles, California
90095, USA\\
 
}
\end{center}
 \vskip 3cm
\noindent
{\Large \bf Abstract} Stochastic processes with multiplicative noise
have been studied independently in several different contexts over the past decades.
We focus on the regime, found for a generic set of control parameters, in which
stochastic processes with multiplicative noise produce intermittency of a special
kind, characterized by a power law probability density distribution. 
We present a review of applications, 
highlight the common physical mechanism and summarize the main known results.
The distribution and statistical properties of the duration of intermittent bursts are 
also characterized in details.

\pagebreak
 
\section{Introduction}
 
Consider the number of fish $X_t$ in a lake in the
$t$-th year. Let $X_{t+1}$ be
related to the population $X_t$ through
\be
X_{t+1} = a X_t + b, \hspace{0.5in}     
\label{eq1}
\ee
where $a = a(t) >0$ and $b=b(t) >0$ are drawn from some probability density function (pdf).
The growth rate $a$ depends on the rate of reproduction and the depletion
rate due to fishing or predation, as well as on environmental conditions, and
is therefore a variable quantity. The quantity $b$ describes the input due to
restocking from an external source such as a fish hatchery in
artificial cases, or from migration from adjoining reservoirs in
natural cases; $b$ can
but need not be constant. 
At first sight, it seems that the linear model (\ref{eq1}) is so
simple that it does not deserve a careful theoretical investigation. However it turns
out that this is not the case\,: see, for example, a rather complicated
mathematical analysis of the problem in \cite{Kesten}. It turns out that model (\ref{eq1})
exhibits an unusual type of intermittency with a power law pdf
 of the variable $X_t$, for a large range of pdf's for $a$ and 
$b$. As we show, the non-trivial properties of this
simple model (\ref{eq1}) come from the competition between the
multiplicative and additive terms. 

If $b = 0$ then (\ref{eq1}) describes a
simple multiplicative process with a distribution which is log-normal
in its central part. If $a=0$, then
$X_t = b$. If $(a(t), b(t))$ are the positive
constants
$(a, b)$, then $X_t$ converges in the limit of large $t$ to the attractor
at
the stable fixed point ${b} \over {1 - a}$ for $a<1$ and diverges if $a >
1$.
In the continuum limit of large t, assuming differentiability, the
difference equation becomes a first order differential equation whose solution is
$X(t) = {{b} \over {1-a}} + C e^{(a - 1) t}$.
The convergence (divergence)
to (from) the fixed point is exponential with the rate $(a - 1)^{-1}$ per unit time.
 
In the case of divergence, limits on growth have often been introduced
through the mechanism of nonlinear saturation, in the absence of external
restocking.  Probably the most studied case of  such nonlinearity is
the logistic equation \cite{May,Ver}, which exhibits a panoply of dynamical
behavior that ranges from bifurcations via a period doubling sequence, to
deterministic chaos, among other properties, as well as certain
universal properties \cite{Collet}.  However, because of the saturation, the
nonlinear dynamics is not able to simulate the observations of
self-similar behavior that extend to large values of the variable
$X(t)$. 

Although the model (\ref{eq1}) appears to be an
apparently innocuous AR(1) (autoregressive of order 1) process, the {\it
random and positive} multiplicative coefficient $a$ lead
to non-trivial intermittent behavior for a large class of distributions
for $a$. Standard stochastic
population models \cite{Bartlett} are often described by (\ref{eq1}) with {\it
fixed} coefficients. The intermittency arises because of
the multiplicative noise structure, taken together with the additive source, and
are not a property of usual AR processes with fixed coefficients \cite{AR}.
Earlier investigations of auto-regressive models
with random coefficients \cite{Nich}
have focused on values of the parameters wherein $X_t$ exhibits a
stationary variance, and thus exclude the highly intermittent regime
we describe here which is characterized by power law distributions 
with finite and infinite variance.
 
The formal solution of (\ref{eq1}) for $t \ge 1$ can be obtained explicitly
\be
X_{t} = (\prod_{l=0}^{N-1} a(l)) X_0 + \sum_{l=0}^{N-1} b(l)
\prod_{m=l+1}^{N-1} a(m)~~,
\label{azjhg}
\ee
where, to deal with $l=N-1$, we define $\prod_{m=N}^{N-1} a(m) \equiv 1$. 
Because of the successive multiplicative operations on $a$ in the
iteration of $X_t$, it is the average logarithmic
growth rate $\langle \log a \rangle$ that controls whether the
population grows or dies. If the average logarithmic growth rate $\langle \log
a \rangle$ is negative due to overfishing and other
environmental influences (in the
example), then the population must ultimately
disappear if there is no restocking through the $b$ term. On the
other hand, $X_t$ diverges with increasing time if $\langle \log a
\rangle \geq 0$. Thus,
the process decays or grows depending on whether the rate $t_c^{-1} =
\langle \log a \rangle$ is negative or positive. In the case of decay, the
introduction of new stock $b$ has a persistent influence on $X_t$ over the
correlation time $|t_c|$. In absence of nonlinear saturation mechanisms, 
we consider first the regime where $\langle \log a \rangle$ is negative.
From (2), it is clear that the Lyapunov exponent is nothing but
$\langle \log a \rangle$. Since most of the interesting regime occurs
for negative $\langle \log a \rangle$, 
this would give again the impression of 
a trivial behavior, if the additive term was not present. 

Model (\ref{eq1}) has similarities with the on-off intermittency described in \cite{onoff}.
In particular, the addition of additive noise on top of the multiplicative
random forcing of the bifurcation parameter can be described by a biased random walk
repelled from the origin in both problems \cite{onoff,Multi}. The same 
mechanism of intermittent multiplicative amplifications applies to both problems.
However, the 
main difference is that the nonlinear saturation in the models studied in 
\cite{onoff} prevents the appearence of large excursions and the power law tail 
does not exist. Most of the analysis of on-off intermittency has been devoted to 
the unstable regime, corresponding in our notations to $\langle \log a \rangle > 0$,
and to the calculation of the distribution of probability for a laminar phase of a
given length. In contrast, we stress that model (\ref{eq1}) has its most 
interesting regime for the ``stable'' case  $\langle \log a \rangle < 0$ and we
discuss the pdf of the variable itself.

Ref.(\cite{Multi}) presents figures of typical
evolutionary sequence $X_t$ for the uniform distributions $0.48 \leq a
\leq 1.48$ and $0 \leq b \leq 1$. In this case $\langle \log a \rangle
= -0.06747$ and $\langle a \rangle = 0.98$. For these parameters,
$\langle X_t \rangle = {\langle b \rangle \over 1 - \langle a \rangle}
= 25$. Most of the time, $X_t$ is significantly less than its average,
while rare intermittent bursts propel it to very large values.
Qualitatively, this behavior arises because of the occurrence of
sequences of successive individual values of $a$ that are larger than
$1$ which, although rare, are always bound to occur in this random
process. The persistence of the temporal behavior has a decay
which is $\exp[\langle \log a \rangle \tau ]$ on the average,
with correlation time
$1/|\langle \log a \rangle| = 14.8$ of the influence of
restocking by the amount $b(t-\tau)$ at a time $\tau$ in the past on the
present value $X_t$. The distribution of $X_t$ from a numerical
realization with the properties above gives an histogram characterized by a power law
tail 
\be
P(X) \sim X^{-(1+\mu)}  ~~~ ,  
\label{eqdfs}
\ee
and a rolloff at smaller values of $X_t$, which is a result that is mandated by
the fact that as $X(t) \rightarrow 0$ the process is dominated by the
injection of new stock $0 < b < 1$, so that the population is repelled from a zero value.
 
In the next section, we discuss a series of
 applications such as population dynamics with external sources,
epidemics where we provide a rationalization for observed power law data, finance and insurance,
immigration and investment portfolios where we generalize (\ref{eq1}) to expanding processes, and
the internet. In section 3, we review and synthesize useful results. First, we recall 
 a generalization of equation (\ref{eq1}) to general 
contractive maps with intermittent bursts with a repulsion from the origin. We then explain how
the power law pdf can be calculated. We connect these processes to auto-catalytic equations that were
found to also exhibit a power law pdf (for the same reason),
but whose treatment using the Fokker-Planck formalism was limited to
gaussian (multiplicative) noise. We briefly mention log-periodicity that was recently found
as corrections to the leading power law pdf. We exhibit the relationship with intermittency
in nonlinear dynamical system and discuss the multiaffine structure of the corresponding
time series. We end this section by reviewing and extending known results on the durations
of the intermittent bursts and their distribution. We also summarize our knowledge on the 
extremes of the random variable $X_t$. Section 4 proposes a nonlinear extension to the process
(\ref{eq1})
that includes the optimization of the restocking term $b$ to develop an optimal strategy for 
population control. Section 5 concludes.

\section{Domains of application}

\subsection{Population dynamics with external sources}

Equation (\ref{eq1}) 
is the simplest discrete map with external source one can think of. Notwithstanding the
simplicity of its formulation, it exhibits a rich phenomenology, suitable to 
apply to population dynamics. Beyond the example of the fish population with restocking,
agriculture provides another example, in which one wishes to model the
annual
fluctuations of the size of a crop in the case where, on average, the crop
that is harvested leaves a seed crop residue that is
less than that needed to fully reseed for the following year; in this case the
variable $b$ represents the seed purchased or otherwise obtained
from outside sources.

Similarly, the large biological variability of phytoplankton blooms in
shallow
coastal ecosystems, such as estuaries, lagoons, bays and tidal rivers,
can also be explained by a population budget given by equation (\ref{eq1}) \cite{Cloern}
where $a$  accounts for the difference between growth rate and loss rate
to respiration, to pelagic and benthic grazing, and
to exchanges of biomass vertically between the top of the sediments and
the overlying water column and horizontally due to advective and diffusion
transport; the $b$ term models the effects of injection of
new individuals across the boundaries of a given ecosystem.

The results for (\ref{eq1}) might explain the spontaneous large
variability
of  such systems that are observed.
Among still other problems, it would be interesting
to explore whether the variability of long term geophysical
time series, such as river discharge or rainfall, which have been
proposed in \cite{Man} to be self-affine (see (\ref{eq14})), can be 
framed in these simple terms or within their multi-dimensional
generalizations described below.

In population dynamics, empirical observations show that
adjacent generations tend to be less different than distant ones, a problem
that has plagued standard nonlinear dynamical models \cite{Cohen,Sugihara}.
Models (\ref{eq1}) and (\ref{eq2}) do not share this difficulty and agree
with this empirical fact. This suggests that intermittent
multiplicative noise might be a useful ingredient for future modelling of 
biological populations.

\subsection{Epidemics}
 
The number $S$ of new cases of people infected with
measles in isolated islands in the North Atlantic per epidemic event
is observed to have a power law distribution with exponent about $0.3$ (called 
$\beta$ in \cite{Rhodes})\,:
\be
P(S) dS \sim S^{-(1+\beta)} ~dS~.
\ee
The durations of epidemic
events also have a power law distribution $P(\tau)$ with an exponent
$\gamma \approx 0.8$. Equation (\ref{eq1}) is a simple model of multiplicative growth
that accounts for the observation of the power law distributions.
The multiplicative term in (\ref{eq1}) represents the
spread though personal infectious contact and the additive
term is representative of the introduction of new strains by
agents from outside. Using this insight,
we can predict $\gamma=1$ from the following self-consistent argument.
 
Let $X_t$ be the number
of new cases between $t-1$ and $t$ in (\ref{eq1}) which is distributed according to
a power law distribution $P(X) \sim X^{-(1+\mu)}$, as given in (\ref{eqdfs}) and (\ref{eq6}), with
exponent $\mu$. Notice that $\mu$ is a priori different from $\beta$ as $\mu$ describes the pdf of
fluctuations
within an epidemic event while $\beta$ describes the pdf of the total size of that event.
The total number of cases in an epidemic event of duration
$\tau$ is 
\be
S_{\tau} = \sum_0^{\tau} X_t~.
\ee
 For $\mu < 1$, we find  
 \be
 S_{\tau} \sim \tau^{1/\mu}~.
 \ee
  From the conservation of probability under a
change
of variable, we get 
\be
P(\tau) d\tau \sim \tau^{-(1+\beta/\mu)}~d\tau~,
\ee
 {\it i.e.} 
 \be
 \gamma \equiv {\beta \over \mu}~.
 \ee
 In section 3.8 below, eq.(\ref{exprejjf}) finds, from a random walk argument,
  that $\gamma = 1/2$ in the uncorrelated case. If we include correlations by an
  argument of self-similarity of epidemics within epidemics, i.e.
the burst of new cases within a given epidemic event is a small epidemic
event in itself (we define an epidemic event to be
the sudden awakening from a relatively
dormant state),  we get $\beta = \mu$ and thus
$\gamma=1$. This seems consistent with the observation
$\gamma \approx 0.8$ within statistical estimates of the error in
the exponent in power law
distributions \cite{Rank}.

\subsection{Finance and insurance applications with relation to ARCH(1) process}
 
 The relation between (\ref{eq1}) and finance has been pointed out in several papers (see for instance
 \cite{Embr}. Suppose you invest $b(t)$ at time 0,1,2,... in a bond with interest 
 time-dependent interest 
 rate $r_t$. The accumulated value 
 $X_t$ at time $t$ of the interest payments made at times 0,..., t, assuming $Y_0= 1$ is given
 by the equation
 \be
 X_{t+1} = b(t) + r_t X_t~.
 \ee
 The inverse problem to accumulation is discounting. Suppose payments $b(i)$ are made at times 
 $0, 1, 2,...$. Given the interest rate $r_i$ between time $i$ and $i+1$, the discounted
 value at time $0$ of all those payments made till time $t$ is 
 $$
 b()0 + b(1) (1+r_{t-1})^{-1} + b(2) (1+r_{t-1})^{-1} (1+r_{t-2})^{-1} + ...
 + b(t) (1+r_{t-1})^{-1} ... (1+r_{0})^{-1}~.
 $$
 This has exactly the same form as (\ref{azjhg}) and obeys the same equation.
 
 The stochastic difference equation (\ref{eq1}) is also obtained from the ARCH(1)
 (autoregressive-conditionally-heteroscedastic) models of log-returns\,:
 \be
 R_{t+1} = \sqrt{b + a R_t^2} ~Z_t~,
 \label{arch}
 \ee
 where $Z_t$ is a gaussian random variable of zero mean and unit variance. This process (\ref{arch})
 describes a persistence and thus clustering of volatilities $R_t^2$. Indeed, the factor
 $(b + a R_t^2)^{1 \over 2}$ ensures that the amplitude of the motion $R_{t+1}$ is controlled by
 the past realization of the amplitude $R_t^2$. Now, calling $X_t \equiv \langle R_t^2 \rangle$, 
 where the average is carried out over the realization of $Z_t$, 
 it is clear that
 (\ref{arch}) is equivalent to (\ref{eq1}), if we allow the coefficient $a$ and $b$ to depend on
 time (independently from $Z_t$).
 
 In the context of insurance, $X_t$ can be interpreted as the value of a perpetuity\,: the payments
 $b(t)$ are made at the beginning of each period and the accumulated payments $X_{t-1}$ are subject
 to interest. The name perpetuity comes from ``perpetual payment streams''. See \cite{Gerber} for an 
 introduction to perpetuities from a life insurance point of view and \cite{Dufresne} from a pension 
 fond point of view.

\subsection{Immigration and investment portfolios}

\subsubsection{Expanding regime}
 
Consider the case $\langle \log a \rangle > 0$ and define
\be
r \equiv lim_{t \to \infty} {1 \over t} \langle \log b \rangle~.
\ee
If $r = 0$, the additive term $b$ becomes unimportant
since $X_t$ now diverges asymptotically
and the process becomes purely multiplicative.
We restrict our attention to the case where $r$ is
strictly positive. If
$\langle \log a \rangle > r$, $X_t$ diverges exponentially with
log-normal fluctuations, since again $X_t$ becomes asymptotically the
result of a
purely multiplicative process. A novel regime occurs when $\langle \log a
\rangle < r$.
In this case, let 
\be
b(t) = e^{r(t+1)} {\hat b}(t)~,
\ee
 where ${\hat b}(t)$ is a
stochastic variable of order one. Let 
\be
a(t) = e^{r}~{\hat a}(t) ~.
\ee
 Since $r > \langle \log a \rangle$, then $\langle \log
{\hat a} \rangle < 0$. Finally let  
\be
X_t = e^{rt}  {\hat X}_t~.
\ee
Equation (\ref{eq1}) is then transformed into 
\be
{\hat X}_{t+1} = {\hat a} {\hat X}_t + {\hat b}~,
\ee
 where ${\hat a} $ and ${\hat b}$ obey the
conditions of our previous analysis exactly. Thus ${\hat X}_{t}$ has a
well-defined asymptotic non-singular pdf with a power law tail for
large values.
 
The sequence $X_t$ is not stationary since it blows up
exponentially at the average rate $r$.  But for a given $t$, one can
characterize
fully the {\it local} distribution of $X$ over a small time interval from the
asymptotic pdf of  ${\hat X}_{t}$. Around the average exponential
growth
at rate $r$, $X_t$ exhibits large fluctuations. Locally in time, the
distribution
of $X_t$ is of the form 
\be
P(X_t) \sim {e^{\mu rt} X_t^{-(1+\mu)}}~,
\ee
where $\mu$ is a solution of 
\be
\langle a^{\mu} \rangle = e^{r\mu}~.
\ee
The simulation shown in Figure 3 shows an average exponential
growth
with superimposed fluctuations, much as in the case of Fig. 1.
 
There is a parallel situation to that considered above
if $\langle \log a \rangle > 0$ with the constraint that $X_t$ be smaller
than some threshold $X_c>0$. The pdf
of $X_t$ is then proportional to $X_t^{\mu -1}$ for $0 \leq X_t < X_c$,
with $\mu$ given by equation (7) as before. This problem is the same as
that studied by Grinstein et al. \cite{Grinstein} for higher dimensional
versions of the model.
 
 \subsubsection{Growth of immigrant populations}
 
As an example of exponential growth, consider the problem of
the growth of immigrant populations. Assume that the
 immigration flow $b$ is increasing at a 
rate $r$.  The national population may or may not
grow exponentially with an average rate $\langle \log a \rangle$ which can
be negative (as for Germany in the last quarter of century) or positive but with
a growth rate smaller than that of the immigrant population.
 This condition holds in general
for several occidental countries that are magnets for immigration
from the developing countries.
For example, the increased rate of flow $b$ might
be due to an average growth
rate of population outside the country which is larger than the
growth rate within the country itself, although we do not have to
restrict ourselves to this case.   From the analysis above, we conclude
that, due to the influx of immigrants from poorer countries with
rapidly
growing populations, the growth rate of the host country ultimately
approaches that of the poorer countries that are the sources of the
influx. Our model predicts self-similar fluctuations of the
host-country
population around the average growth rate, due to the interplay
between the natural fluctuations of the intrinsic growth rate of the
host
country and the immigrant flux. This model assumes that immigrants will
abandon the fertility rate they had previously and
adjust it to that of the host country, on a time scale
of the fluctuations of the intrinsic growth rate {\footnote{Indeed, at the next
time step, the immigrant term $b$ becomes part of the total population $X$
which is updated with the (reduced) national growth rate $a$.}}.
We have not tested
these predictions for countries that have had a significant influx of
immigrants. The simple model (\ref{eq1}) is not expected to account for all
growth histories
but it might help to
distinguish between different causes for the growth
of population. We note also that economics Nobel laureate Simon had similar
ideas to explain the growth laws for cities \cite{Simon}.

  \subsubsection{Growth of investment porfolios}
 
Consider the growth of a portfolio of
investments \cite{Soros} such as that of a mutual
fund. Consider a successful fund that
exhibits a higher than market average return. Investors are thereby attracted
to it and may invest new capital at a rate faster than the return rate
itself. Not infrequently, new capital may flow in, even in the presence of
relatively
poor performance \cite{Gruber} {\footnote{This is a zeroth-order
approximation since it can be
expected that correlations exist between the return of
a fund and the confidence of the investors.}}. Thus
the condition $\langle \log a
\rangle < {1 \over t} \langle \log b \rangle$ holds. The growth of the fund
thus occurs not only through reinvestment of returns,
but also through periodic additions under
an investment plan. Due to the stochastic character of returns, the fund
will show regular exponential growth, upon which is superimposed a series of
spurts and gradual decays back to the exponential growth \cite{Soros}; the
fluctuations are
distributed according to a power law. We have assumed that mutual funds are
bought
and sold as net asset value and their liquidity
is linked with their growth. Since liquidity is commonly observed
to be correlated
with the volume of transactions, which is itself correlated with price
volatility \cite{vol}, these considerations might also be relevant to
explain fat tails for price variation pdf's \cite{Haan,Stanley}.
 
Equation (\ref{eq1}) may also describe the evolution of composite capital growth
on the large scale of an
economy \cite{Black}, where \{$a = 1 +$ return $-$ consumption\}
and $b$
is the addition of new
wealth from technological innovations or discoveries
of mineral resources.

 \subsection{The internet}
 
 There are probably many other areas where application of these ideas may prove to be
fruitful. A recent paper by Huberman and Lukose \cite{huber} considers
a multiplicative process to describe the population of users of the internet
which cooperate. The model predicts intermittent congestions with short-lived
spikes but lacks a random source of incoming users (the $b$ term) which leads them
to predict a log-normal distribution for Internet latenties. The reported data
seems to exhibit a tail fatter than lognormal (see their figure 3),
which we are tempted to interpret as the
consequence of the multiplicative process, with intermittent amplification, coupled to
a random additive source. In fact, Takayasu et al. \cite{Takayasu} have found 
a power law pdf for the frequency of jams in Internet traffics.

 \section{Statistical properties}
 
 \subsection{Generalization}
 
 Eq.(\ref{eq1}) defines a stationary process if $\langle \ln b(t) \rangle < 0$.
In order to get a power law pdf, $a(t)$ must sometimes take values larger than one,
corresponding to intermittent amplifications. 
This is not enough\,: the presence of the additive term $b(t)$ (which can
be constant or stochastic) is needed to ensure a ``reinjection'' to finite values, 
susceptible to the intermittent amplifications. It was thus shown \cite{Multi}
that (\ref{eq1}) is only one among many convergent ($\langle \ln b(t) \rangle < 0$)
multiplicative processes with 
repulsion from the origin (due to the $b(t)$ term in (\ref{eq1}))\,:
\be
X_{t+1} = e^{f(X_t, \{a, b,...\})} ~a ~X_t ~.
\label{eq2}
\ee
$f$ has the following properties\,:
\be
f(X_t, \{a, b,...\}) \to 0 ~~~{\rm for} X_t \to \infty~, 
\label{eq3}
\ee
i.e. $X_t$ is a pure multiplicative process when it is large,
\be
f(X_t, \{a,b,...\})) \to \infty ~~~{\rm for} X_t \to 0~, 
\label{eq4}
\ee
i.e. $X_t$ is repelled from $0$. Additional conditions of 
monotonicity and regularity must be added to ensure that the pdf
of $X_t$ is unbound at large $X_t$.

All these processes share the
same power law tail for their pdf $P(x) \sim x^{-1-\mu}$ for large $x$ with $\mu$ solution
of $\langle a(t)^{\mu} \rangle = 1$. $\ln x(t)$ undergoes a random walk with
drift to the left which is repelled from $-\infty$. A simple Boltzmann argument
shows that the stationary 
concentration profile is exponential, leading to the power law pdf in the $x(t)$
variable.

The deterministic version of equation (\ref{eq2})
has been discussed frequently in the biological literature (see
\cite{May} and references therein).
The model (\ref{eq1}) is the special case
$f(X_t, \{a(t), b(t),...\}) = \log (1 + {b(t) \over a(t) X_t})$.
Thus the dynamics of $X_t$ is the result of a multiplicative process
that contracts on average.
However the restocking term $b$
prevents $X_t$ from approaching zero by repulsion from the origin, and
allows the distribution to
converge to a non-trivial asymptotic pdf that turns out to be a power
law under some mild assumptions stated below. In other words, the
restocking term
$b$ which is a repulsion from the origin, corresponds to a
resetting of the dynamics away from $X_t = 0$. The multiplicative
process, with an $a$ that can sometimes have values that are larger than
$1$, ensures
an intermittent sensitive dependence on prior values of $X_t$.
It is noteworthy that these ingredients of
sensitive dependence on prior conditions and reinjection (of new stock),
are also
the two fundamental properties of systems that exhibit chaotic behavior 
\cite{Chaos}.
 
 \subsection{Power law probability density function}
 
Qualitatively, the existence of a limiting distribution for $X_t$
that obeys (\ref{eq2}), for a large class of $f(X, \{a, b,...\})$ as in 
(\ref{eq3},\ref{eq4}), is
guaranteed by the competition between the convergence of $X$ to zero and the
sharp repulsion from it. Kesten's result for model (\ref{eq1}) proves only the 
existence of a power law tail for the pdf but gives no information about
the central part of the pdf, which is assumed at least to be an
integral function. The most general mathematical results for this
class of models are valid only for linear systems (\ref{eq1}) and under 
a number of assumptions about the random coefficients 

In the case of the generalized model, we can also give only a result for
the tail of the pdf. With the additional condition $\partial f(X, \{a,
b,...\})/\partial x \to 0$ for $X \to \infty$, which is satisfied for a
large
subclass of smooth functions that satisfy conditions (\ref{eq3},\ref{eq4}), 
the pdf of
$X$
is the solution of
\be
pdf\biggl(X e^{-f(X, \{\lambda, b,...\})}\biggl) ~=~ pdf\biggl( a X \biggl)~~~.  
\label{eq5}
\ee
The expression (\ref{eq5}) means that
the l.h.s. and r.h.s. have the same distribution.
The solution to this problem has been given in \cite{Multi} as a solution to
a Wiener-Hopf equation for the tail.
All the models (\ref{eq2}) are characterized by a
pdf with a tail that decays asymptotically for large $X$ as a power
law
\be
P(X) \sim X^{-(1+\mu)}  ~~~ ,  
\label{eq6}
\ee
if there is a solution  $\mu > 0$ {\footnote{In general, when there is a
real
positive solution, there are also an infinitely discrete number of complex
solutions (see below for their interpretation).}}
of the equation
\be
\langle a^{\mu} \rangle = 1~~~~. 
\label{eq7}
\ee

It has been shown in \cite{Jogi} that the exponent $\mu$ 
does not depend on the realizations of
$b$ if the distribution of $a$ is smooth, and more generally, 
on the specific form of the repulsion from zero.
For the example shown in the figures, the
numerical solution to (\ref{eq7}) is $\mu \approx 1.467$. The
pdf $P(X)$ is always integrable at $X \to \infty$ because $\mu > 0$. Its
integrability for small $X$ is ensured automatically if $b$ is bounded from
below for $b > 0$ and can be shown to be true even when the pdf of $b$
extends down to zero, provided it is not too singular at $0$.
 
The result (\ref{eq6}) with property (\ref{eq7}) was proved first for
model (\ref{eq1}) by Kesten \cite{Kesten}, and was then revisited by several authors
in the differing contexts of finance \cite{Haan} and 1D random-field Ising
models \cite{Calan}.
Recently, Levy, Solomon and Ram \cite{Solomon} have rederived this result for a
different submodel of the class (\ref{eq2})
 by the use of the extremal properties of the $G -${\it
harmonic} functions on non-compact groups
\cite{Choquet}, which are translational groups in this paper. In \cite{Multi},
a mapping to a biased random walk recovers these results. In \cite{Taka},
the method of characteristic functions has been used to recover the power law
for the regime of infinite variance corresponding to $\mu \leq 2$.

As a consequence of the relatively short-range correlations of $X_t$ and the
properties of linear combinations of variables with power law pdf
tails, the pdf of the variations $(X_{t+1} - X_t),$ and of higher order
differences as well, also have a tail of the form (\ref{eq11}) with the same
exponent $\mu$.
 
In the limit of small disorder on the multiplicative
noise $\langle (\log a)^2 \rangle - \langle \log a \rangle^2 \ll 1$, we can
solve (\ref{eq7}) by an expansion in $\langle a^{\mu} \rangle$ to second order in
the cumulants of the pdf of $a$ and get the approximation
\be
\mu \approx {{|\langle \log
a \rangle|} \over {\langle (\log a)^2 \rangle - \langle \log a
\rangle^2}}  ~. 
\label{eq8}
\ee
This approximation is accurate for narrow pdf's
and is exact if the pdf of $\log a$ is gaussian.  For the case of
the uniform distribution of the example, the estimate (\ref{eq8}) is $\mu
\approx 1.359$, which differs from the exact value $1.467$ by about
$7\%$.
 
 \subsection{Relation with auto-catalytic stochastic ODE}
 
In the context of auto-catalytic equations that lead to multiplicative
stochastic equations, Graham and Schenzle \cite{Graham} have determined the
exact asymptotic pdf for the variable $X$ obeying the equation
\be
{dX \over dt} = - r X + p X^{1-p} + \eta X ~~~~, 
\label{eq9}
\ee
where $r$ is the decay rate in the absence of the last two
terms and $\eta$ is a multiplicative {\it Gaussian} noise with zero mean.
If
$p>0$, then $X^{1-p}$ is negligible for large $X$ compared to $r X$
but dominates as $X \to 0$. This terms plays exactly the same role
as the function $f$ in the discrete equation (\ref{eq2}) by guaranteeing
repulsion from $X = 0$. It is clear that a discretization of (\ref{eq9})
 yields (\ref{eq2}) with
$a(t) = 1-r + \eta(t)$ {\footnote{Care must be taken to choose
the correct representation (Ito or Stratonovich) and use the differential
calculus accordingly in going from the discrete to the continuous
description \cite{Kampen}.}}. In the case of
Gaussian noise which is the only one that has been studied thus far,
expression (\ref{eq6}) is recovered, and the exponent is found again
to be
completely independent of the specifics of the repulsion from the origin,
which is
parameterized by the exponent $p>0$ {\footnote{For $p<0$, the repulsion is
from infinity rather that from zero, and the dynamics is
completely different.}} in this case. It is of interest to note that the discrete
systems (\ref{eq1}) and (\ref{eq2}) with non-Gaussian noise lead to pdf's belonging to
a {\it
different universality class} than in the case of Gaussian noise, i.e.
these systems will have different
exponents for the same average and variance of the multiplicative noise.
This property is due to the importance of rare large deviations on the
determination of the exponent $\mu$.
 
 \subsection{Generalization to multi-dimensional processes}
 
For completeness, there is a recent generalization of eq.(\ref{eq9}) to
multidimensional systems \cite{Grinstein,Hwa},
\be
{\partial X \over \partial t} = D {\partial^2 X \over \partial x^2}
- r X + p X^{1-p} + \eta X~~~~. 
\label{eq10}
\ee
where $X(x,t)$ is now a field in a $d-$dimensional space $x$ and $D$ is a
diffusion coefficient. Qualitatively, this equation describes a
$d-$continuous
infinity of variables $X$, each of which follows a multiplicative
stochastic dynamics having the forms (\ref{eq2}) or (\ref{eq9}), 
and in the discretized
equivalent is coupled to
nearest neighbors through a diffusion term.
The problems (\ref{eq9}) and (\ref{eq2}) correspond to the {\it
zero}-dimensional case. Grinstein et al. \cite{Grinstein} have studied the
case $p<0$ in an arbitrary number of dimensions, which leads to a repulsion from
$+\infty$. Thus
this case cannot lead to a pdf with a long tail and instead must belong to a
completely
different regime. Munoz and Hwa \cite{Hwa}
have considered the case $p>0$ which is relevant to our
discussion, and also find a power law decay for the pdf of $X$. However, the
determination of the exponent is done by numerical simulations and there
are
no exact results available for $d>0$.
 
 \subsection{Complex exponents and log-periodicity}
 
For $a$'s with not-too-wide pdf's, complex solutions of (\ref{eq7})
for the exponent $\mu$ can also be found. Complex values of $\mu$ lead
to log-periodic oscillations that are superimposed on
the lowest-order behavior which is power law \cite{log1,log2}, as can be seen in
the tail of figure 2. The underlying discrete scale invariance \cite{Houches} has
recently been found to be a widespread property of irreversible
out-of-equilibrium processes \cite{example1,example2,example3,example4}.
This log-periodicity overlaying the leading power law behavior of the probability
density distribution
is studied in details in \cite{Jogi} with emphasis on the 
progressive smoothing of the log-periodic structures
as the randomness increases in order to test its robustness.
It is shown that the
log-periodicity is due to the intermittent amplifying multiplicative events.

\subsection{Relation with intermittency in nonlinear dynamical systems}

These results should not give the impression that we have a complete
understanding of the systems (\ref{eq1}) or (\ref{eq2}). The results we have
catalogued  only describe the asymptotic tail for large $X$ in the
infinite time limit. The problem of the rate of convergence is much
more involved and has only been addressed
for eq. (\ref{eq9}) in the case of Gaussian noise \cite{Graham}. Furthermore, the
apparent
simple linearity of (\ref{eq1}) masks equivalent nonlinear
properties that are non-trivial. As mentioned, these systems have
an intermittent
sensitivity to both the prior values of $X$ and the reinjection mechanism. The
analogy
with deterministic dynamical systems is even closer. 
Suppose that $a$
is not truly stochastic but is itself generated by a deterministic dynamical
equation, say $a(t) = \epsilon + r_t$, the latter given by
the familiar logistic map
$r_{t+1} = 4 r_t (1-r_t)$. The logistic map is
well-known to be fully chaotic and has a invariant measure
$P(r) = {1 \over \pi} [r(1-r)]^{-1/2}$. For positive but not too
large $\epsilon$, $a$ obeys the condition $\langle \log a \rangle < 0$ while
occasional individual realizations are larger than $1$. Thus the
fully
deterministic coupled system,
\be
X_{t+1} = (\epsilon + r_t) X_t + b~~, 
\label{eq11}
\ee
\be
r_{t+1} = 4 r_t (1-r_t)~~,
\label{eq12}
\ee
has the properties summarized above. We obtain
the example exactly, if we replace (12) by the tent map
\be
r_{t+1} = 2 r_t ~~~~ mod(2) ~~,   
\label{eq13}
\ee
which is equivalent to (\ref{eq12}) under a change of variable and has a uniform
invariant measure in the interval $0<r<1$. The choice $\epsilon = 0.48$
gives the numerical example discussed in the introduction. The equivalence between
the stochastic and deterministic chaotic processes for similar random
walks has been checked in \cite{Alain}. 
In our present problem, this result is correct as can be proved using the 
transition operator approach, which consists in 
describing the image of an arbitrary distribution density under the action of our 
random process. Ref.\cite{Jogi} then proves that
the representation of the transition 
operator depends only on the stationary probabilities of the random 
couples $(a, b)$, but not on the probabilities of the transitions between 
them, i.e. on the fact that the $r_t$ are correlated.

\subsection{Self-affinity and multiself-affinity}
 
The process $X_t$ defined by (\ref{eq1}) or (\ref{eq2}) is self-affine and can be
characterized by its Hurst exponent $H$ for $1 \leq \mu \leq 2$,
through the expression \cite{Hurst}
\be
\langle (X_{t+\Delta t} - X_t)^2 \rangle^{1
\over 2} \sim (\Delta t)^H , 
\label{eq14}
\ee
with
\be
{1 \over 2} \leq H={1
\over \mu} \leq 1 ~,
\label{eq15}
\ee
For the parameters of figure 1, $H = 0.68$.
The relation (\ref{eq14}) is valid for $\Delta t > |t_c|$.
To see this, the typical largest value $X_{max}$ among $\Delta t$
realizations
of the variable $X_t$ is given by
\be
\Delta t \int_{X_{max}}^{+ \infty} P(X) dX \sim 1 ~~, 
\label{eq16}
\ee
which leads to
\be
X_{max} \sim \Delta t^{1/ \mu} ~~~.  
\label{eq17}
\ee
The second moment $\langle (X_{t+\Delta t} - X_t)^2 \rangle$ is
then given by
\be
\langle (X_{t+\Delta t} - X_t)^2 \rangle \sim
\Delta t \int^{X_{max}} X^2 P(X) d X \sim
\Delta t X_{max}^{2-\mu} \sim \Delta t^{2/ \mu} ~~. 
\label{eq18}
\ee
Equation (\ref{eq15}) follows.
 
For a long sequence, the values $X_t$ exhibit multi-self-affinity,
as are log-normal distributions \cite{Halsey,Redner}.
It is also possible to  add a
space dimension to the model by constructing multiaffine fields so
that the field obeys eq. (\ref{eq1})
locally,
and thus has all its properties. To do this, we use the
construction given by Benzi, et al. \cite{Benzi} with a mother wavelet
basis
function which is a constant over the unit interval; these operations
will be
developed further elsewhere.
The above properties
hold for infinite time sequences. In the practical case of samples of
finite length, finite size effects round off the power tails: for $X_t
> e^{\sqrt{D t}}$, (6) is approximately transformed into the log-normal law
$\exp \biggl[ - {1 \over 2D t} ( \log X_t  -  V t)^2 \biggl]$, where
$V \equiv \langle \log a \rangle$. This result arises because these
large values of $X_t$ are in the ``free'' multiplicative regime and are
remote from the influence of the repulsive, additive terms $b$ that
are important near $X = 0$.

 \subsection{Extremes and durations of the intermittent bursts}
 
 It is useful to characterize further the nature of the intermittent bursts.
 It is possible to quantify the distribution of extremely large bursts. 
 In this goal, we adapt the result given by \cite{emmbre}
 for ARCH(1) processes and translate it to the case of (\ref{eq1}). We thus obtain
 \be
 lim_{t \to \infty} P \biggl( max(X_1, X_2, ..., X_t) \leq x t^{1/\mu} \biggl)
  = \exp [-c \theta x^{-\mu}]~,
  \label{etry}
  \ee
  where the exponent $\mu$ is given by (\ref{eq7}), the constant $c$ is given by
  \be
  c = {\langle (b + a X_t)^{\mu} - (a X_t)^{\mu} \rangle \over \mu \langle 
  a^{\mu} \ln a \rangle}~,
  \ee
  and 
  \be
  \theta = \mu \int_1^{\infty} P\biggl(max_{t \geq 1} \prod_{i=1}^t a(i) \leq {1 \over y} \biggl)
  ~y^{-1-\mu} dy~.
  \label{theoeo}
  \ee
 Expression (\ref{etry}) shows that the stationary process (\ref{eq1}) have similar extremal
 properties as a sequence of independent and identically distributed random variables with the
 same probability density function. In the mathematical literature on extremes \cite{emmbre},
 $\theta$ is known as the ``extremal index''. It quantifies the role of dependence between the 
 successive $X_t$ in the realization of extremal values. In particular, $\theta < 1$ implies
 a smaller achieved extreme compared to the case of iid random variables with the same
 powerlaw distribution.
 
 It is also possible to quantify the subset of times $1 \leq \{t_e\} \leq t$
 at which $X_{t_e}$ exceeds the threshold $x t^{1 \over \mu}$. In other words, 
 among $X_1, X_2, ..., X_t$, some are above $x t^{1 \over \mu}$. What is the process
 describing these times of exceedance? It can be shown \cite{emmbre} that this subset converges in
 distribution to a compound Poisson process with intensity $c \theta x^{-\mu}$
 and cluster probabilities 
 \be
 \pi_k = {\theta_k - \theta_{k+1} \over \theta}~,
 \ee
 where
 \be
 \theta_k = \mu \int_1^{\infty} P\biggl( \rm{card} \biggl[ t / 
 \prod_{i=1}^t a(i) > {1 \over y} \biggl]
 = k-1 \biggl) y^{-1-\mu} dy~.
 \ee
 Note that $\theta_1 = \theta$ as defined by (\ref{theoeo}). $\rm{card}\biggl[ S \biggl]$ 
 stands for the cardinal of
 the set $S$, i.e. the number of elements in that set.
 
 Intuitively, the points of exceedance of a given threshold can be obtained in terms
 of the random walk $S_0=0$, $S_t = \sum_{i=1}^t \ln a(i)$ \cite{emmbre} (page 475).
   This is exact in the process
studied in \cite{Solomon} defined by $X_{t+1} = \sup\{ a(t) X_t, 1 \}$, which defines 
a multiplicative process that is reflected from the left boundary $X=1$. This
reflection ensures the repulsion from the origin described as the 
generic mechanism for the generation of power law pdf's by convergent multiplicative 
processes. Taking the
logarithm and defining $x_t \equiv \ln X_t$, 
$\ln X_{t+1} = \sup \{\ln X_t + \ln a(t), 0 \}$ which defines a random walk
with a bias to the left ($\langle \ln a \rangle < 0$) and reflected at the origin.
In between two successive reflections, this is a pure random walk. The process is thus
the union of pieces of biased random walks, all starting from the origin and returning
to it for the first time. At each collision with the origin, the random walk looses the
memory of past motion. This correspondence allows us to get the exact distribution of 
durations between return to the origin, i.e. the distribution of the peak durations.
Notice that this distribution $F(t,x)$ (which is the probability for the random walker
to reach position $x$ at time $t$, starting from the origin at time $0$) 
also describes the width of a peak above an arbitrarily 
defined threshold, as the probability of return to the origin is independent of the
definition of the origin. This distribution is a power law $\sim t^{-{3 \over 2}}$ 
truncated for large $t$ by the presence of the negative drift 
$v \equiv \langle \ln a \rangle$ which tends to bring back the random walk faster to the
origin. The exact expression is obtained from the technique of generating functions
\cite{Fisher}
and reads (see \cite{meee} for an exact explicit relation and also \cite{mont})
\be
F_v(t,x) = (1 - v^2)^{t \over 2} ~ e^{-\alpha x} F_{v=0}(t,x)~,
\ee
where $\alpha$ is defined by $\cosh \alpha = {1 \over \sqrt{1 - v^2}}$ (leading to $\alpha \to 0$
for $v \to 0$) and $F_{v=0}(t,x)$ is the probability of first passage at $x$ 
starting from the origin at time $0$, under no drift. The factor 
$(1 - v^2)^{t \over 2} ~ e^{-\alpha x}$ thus embodies the effect of the bias
$v \equiv \langle \ln a \rangle <0$.

The durations of intermittent amplifications are thus distributed according to $F_v(t,0)$
giving the pdf of the first return to the origin, starting from the origin at time $0$.
Since $F_{v=0}(t,0) \sim  t^{-{3 \over 2}}$, we get 
\be
F_v(t,0) \sim t^{-{3 \over 2}} ~e^{{\ln (1-v^2) \over 2} ~t} \approx 
t^{-{3 \over 2}} ~e^{-{v^2 \over 2} ~t}~,
\label{exprejjf}
\ee
where the last expression is valid for small drift. The same tail with a power law truncated
by an exponential describes the other processes (\ref{eq2}).

For the example of section 2, 
$v \equiv \langle \ln a \rangle = -0.06747$, leading to a characteristic time 
$|{2 \over \ln (1-v^2)}| \approx 440$. The pdf of the durations of intermittent amplifications
is thus indistinguishable from the power law $t^{-{3 \over 2}}$ for durations less than $440$ and
crosses over to an exponential tail for large durations. Figure 1 shows a sequence which is clearly 
in the first power law regime for the duration pdf.

\section{A non-linear extension\,: optimization of restocking strategy}
 
 Up to now, the restocking term $b(t)$ has been considered to be independent of $X(t)$. As a first
 exercise, we now examine a case where $b(t)$ becomes a function of $X(t-1)$ so
 as to optimize a restocking strategy. This problem is motivated by the 
 observation that, whether we consider the general problems of game management or the
specific example of fish restocking,
large fluctuations in population size from year to year are
to be expected. One could thus hope to cushion the fluctuations by exerting
a control on $b(t)$. In this spirit, 
variations of the model (\ref{eq1}) have recently been
proposed for the analysis of crop control in the presence of weed infestion 
\cite{Hugues}, in which the control of restocking is done on an action on 
the multiplicative term. Here, we analyze the restocking
strategies defined by the action of $b$, where $b$ can become 
a function of $X_t$ that do not modify the
power-law property of the tail,  for a broad class of functions
$b(t,X_t)$ {\footnote{We note that the linear relation $b = \alpha X_t$
does not
preserve the power-law property
since this yields the purely multiplicative process
$X_{t+1} = (a(t) + \alpha) X_t$. More generally, if $b$ vanishes
with $X_t$, the influence of restocking disappears for small $X$.  It
is the mechanism of repulsion from the origin induced by the
$b$ term that leads to the power law distribution.}}, if the
growth
rate $a$ is unknown and fluctuating. Indeed,
large scale fluctuations are a robust feature of the intermittent
multiplicative process (\ref{eq1}) which do not depend on the specific
nature of the reinjection mechanism at small $X(t)$. Optimized strategies $b(t)$ can
be defined, even in the absence of correlations in the time
series $a(t)$, if one knows the pdf $P_a(a)$. 

We refer to \cite{Weissbuch} for an effort to model
explicitly  the dynamics of the fishermen
behavior, coupled to the ecological and economic dynamics, which is partly based on
neural nets. We also neglect nonlinear corrections between the annual number of 
offsprings and the size of the stock, which is important in absence of
restocking \cite{Cook} but is of minor concern when $\langle \log a \rangle < 0$ for
which the restocking term is dominating the dynamics.

Suppose we want to prevent the
population $X_t$ from decreasing below some minimum $X_{min}$, and that any
restocking action has a price, which can be taken to be a constant
(overhead) plus a term that is proportional to the amount of added
game, to a first
approximation.
The mathematical solution of this problem demands the definition of
a cost function to be minimized. Many choices are possible and can
be treated similarly. For the sake of
illustration, let the cost function be proportional to the
probability
that $X_t$ will be less than some $X_{min}$;  if $X_t < X_{min}$ we
declare that the year
will be  lost for fishermen or hunters and hence also for
the various suppliers of their equipment. In term of economic loss, our
simple model assumes that the price of  such an event is some aggregate
loss
multiplied by the probability that it occurs. The price of restocking
must be added to the price of the event and the sum minimized, which
corresponds to a trade-off. We thus minimize $\int_0^{X_{min}}
P(X_{t+1}|X_t,b(t)) dX_{t+1}  + \lambda b(t)$, with respect to $b(t)$ to
find
the best restocking strategy. The quantity $P(X_{t+1}|X_t,b(t))$
is the pdf of $X_{t+1}$, given the population $X_t$ of the previous
year
and assuming $b(t)$ determined by the optimization. The factor
$\lambda$ is a measure of the price of
restocking
relative to the loss experienced when the population is too small. From (\ref{eq1}),
$P(X_{t+1}|X_t,b(t))$ is simply deduced from $P_a(a)$ by a
change of variable.
The solution of the minimization problem is
\be
P_a({X_{min} - b(t) \over X_t}) = \lambda X_t  .
\label{eq19}
\ee
This equation has a non-zero solution for $b$ only if $\lambda X_t$ is
smaller
than the maximum of $P_a$. If not, the solution is $b = 0$
which
corresponds to the situation where either the previous population $X_t$ was
so
large that next year is almost certain to be a good year without external
action, or the price to restock $\lambda$
is too large. If $\lambda X_t$
is
smaller than the maximum of $P_a$, and $P_a$ has a bell shape,
there
are two solutions to (\ref{eq19}), but only the one which makes
${X_{min} - b(t) \over X_t}$ the least is the genuine minimum
of the cost function, the other being a maximum. The specific optimal
strategy for choosing $b$ thus depends on the detailed
shape of the pdf $P_a$: this
does
not come as a surprise since
the multiplicative process $a$ is the dominant factor in the creation of
stochasticity. We find, in agreement with intuition, that $b$
is larger for smaller $\lambda X_t$. In general, $b$ is a
non-linear function of $X_t$. However, since it goes to zero for large
$X_t$,
we conclude that the fluctuations for large $X_t$ are not modified by
the restocking strategy, which after all is only important for buffering
the low values.

\section{Concluding remarks}
 
We have reviewed the main know properties of intermittent multiplicative processes.
We have shown that they provide a
robust and general model of intermittent self-similar dynamical
processes with power law pdf tails. Even in the {\it complete absence} of
non-linearity, the population $X_t$
exhibits a
non-trivial intermittent dynamics for $\langle \log a \rangle < {1
\over t} \langle \log b \rangle$, whether $b$ be
determined deterministically  or stochastically. If $b$ is bounded
or grows sub-exponentially, then ${1 \over t} \langle \log b \rangle \to
0$
and the multiplicative process has to be convergent  ($\langle \log a
\rangle < 0$). The competition between
the random multiplicative process $a$ and the external source $b$  is
enough to produce intermittency as measured by the Hurst exponent and
power law distributions. These results also hold for the
more general class of models (\ref{eq2}). The
power law distributions are the hallmark of convergent
multiplicative processes repelled from the origin and do not need the fine
tuning of a control parameter as in usual models of criticality \cite{critical}.
The model (\ref{eq1}) provides a simple useful alternative to self-organizing
models with nonlinear interactions \cite{Manne,SOC}. We have however shown how these
stochastic linear systems can
appear to have the properties of a nonlinear dynamical system.

\vskip 1cm
\noindent
{\Large \bf Acknowledgments}
\vskip 0.5cm
I acknowledge useful exchanges with M. Ghil, D. Lettenmaier and C.
Marshall and especially with L. Knopoff. We thank M. Blank, U. Frisch and D. Stauffer
for their remarks that helped improve the content of the paper.
This paper is Publication no. 4710 of the Institute of Geophysics and
Planetary Physics, University of California, Los Angeles.

\end{document}